\documentstyle[11pt,paspfiddle,twoside,epsf]{article}
\def\edcomment#1{\iffalse\marginpar{\raggedright\sl#1\/}\else\relax\fi}
\reversemarginpar

\begin{document}
\title{Studying Structure Formation with the Sloan Digital Sky Survey}
 \author{David H. Weinberg}
\affil{Department of Astronomy, Ohio State University, Columbus, OH 43210, USA}

\begin{abstract}
I review some of the recent results from the SDSS related to galaxies and large
scale structure, including: (1) discovery of coherent, unbound structures in the
stellar halo of the Milky Way, (2) demonstration that the Pal 5 globular cluster
has tidal tails and that the Draco dwarf spheroidal does not, (3) precise 
measurement of the galaxy luminosity function and its variation with galaxy
surface brightness, color, and morphology, (4) detailed examination of the
Fundamental Plane from a sample of 9000 early type galaxies, (5) measurement,
via galaxy-galaxy lensing, of the extended dark matter distributions around
galaxies and their variation with 
galaxy luminosity, morphology, and environment, 
(6) measurements of the galaxy angular power spectrum and of the spatial
correlation function and pairwise velocity 
dispersion as a function of galaxy luminosity and color.  I then turn to a more
abstract discussion of what we can hope to learn, in the long run, from galaxy
clustering in the SDSS and the 2dFGRS. The clustering of a galaxy sample depends
on the mass function and clustering of the dark halo population, and on the Halo
Occupation Distribution (HOD), which specifies the way that galaxies populate 
the halos. Hydrodynamic simulations and semi-analytic models of galaxy formation
make similar predictions for the probability $P(N|M)$ that a halo 
of virial mass $M$ contains $N$ galaxies of a specified type: a non-linear form 
of the mean occupation $N_{\rm avg}(M)$, sub-Poisson fluctuations about the mean
in low mass halos, and a strong dependence of $N_{\rm avg}(M)$ on the age of a 
galaxy's stellar population.  Different galaxy clustering statistics respond to 
different features of the HOD, making it possible to determine the HOD 
empirically given an assumed cosmological model.  Furthermore, changes to 
$\Omega_m$ and/or the linear power spectrum
produce changes in the halo population that would be 
difficult to mask by changing the HOD.  Ultimately, we can hope to have
our cake and eat it too, obtaining strong guidance to the physics of galaxy 
formation by deriving the HOD of different classes of galaxies, while 
simultaneously carrying out precision tests of cosmological models.
\end{abstract}

\section{The SDSS: Goals and Status}

The observational goals of the SDSS have remained stable since its early days:
(1) $u$,$g$,$r$,$i$,$z$ CCD imaging of $10^4\;{\rm deg}^2$ in the 
North Galactic Cap, to a depth $\sim 23\;$mag in the most sensitive bands,
(2) a spectroscopic survey in this area of $10^6$ galaxies, $10^5$ quasars,
and an assortment of stars and other targets, and (3) imaging of three
$2.5^\circ \times 90^\circ$ stripes in the South Galactic Cap, with the
equatorial stripe scanned repeatedly to allow variability studies and
co-added imaging that goes $\sim 2\;$mag deeper than a single scan.
The normal spectroscopic program is carried out on all three stripes in
the south, and additional spectroscopy in the equatorial stripe will
provide deeper samples of quasars and galaxies and more comprehensive
coverage of stellar targets.  
The survey is carried out using a dedicated 2.5-m telescope on Apache Point,
New Mexico, equipped with a mosaic CCD camera (Gunn et al.\ 1998) and two
fiber-fed double spectrographs that can obtain 640 spectra simultaneously
(Uomoto et al., in preparation).
A technical overview of the survey appears
in York et al.\ (2000), and an updated but more focused technical summary
appears in the paper by Stoughton et al.\ (2002) that describes the
Early Data Release.  The quasar survey is reviewed by Schneider et al.
(these proceedings), and the quasar target selection is described by
Richards et al.\ (2002).  There are two samples in the galaxy redshift
survey, a magnitude limited sample to $r=17.77$ comprising 90\% of the
galaxy targets (Strauss et al.\ 2002), and a sparser, deeper sample of
luminous red galaxies (Eisenstein et al.\ 2001). 

After a decade of preparatory work, the SDSS formally began operations on the
auspicious date of 
April 1, 2000.  It is planned to run until summer, 2005.  The total area
covered will depend on the weather between now and then, 
especially the amount of weather that satisfies the seeing and photometric
requirements of the imaging survey.  A reasonable guess is that the northern
survey will cover $\sim 70\%$ of the original $10^4\;{\rm deg}^2$ goal
by summer, 2005.  As of January, 2002, the SDSS had obtained $\sim 3200$
square degrees of imaging and $\sim 230,000$ spectra 
(of which about 80\% are galaxies), including both northern and southern
observations.  The quality of the spectra, which cover the wavelength
range 3800\AA\ to 9200\AA\ at resolution $R\sim 1800$, is spectacular.  
Redshift completeness for spectroscopically observed galaxies is over 99\%, and
for most galaxies the spectra yield stellar velocity dispersions and 
valuable diagnostics of the stellar populations.  The scientific analyses
to date are only scratching the surface of what the spectroscopic data
allow.

In June, 2001, the SDSS released $462\;{\rm deg}^2$ of imaging data 
($\sim 14$ million detected objects) and
$54,000$ follow-up spectra obtained during commissioning observations and
the first phases of the survey proper, as documented by Stoughton et al.\ 
(2002).  In addition to providing data for the larger community, the
Early Data Release is a training exercise for the SDSS collaboration.
One lesson is that releasing data as complex as that in the SDSS
(e.g., radial profiles plus over 80 measured parameters for each photometric 
object, some of which are deblended, some imaged on more than one scan,
some observed spectroscopically) in a useful way is very challenging,
even within the collaboration itself.  The First Data Release will take
place in January, 2003, and subsequent releases will take place on 
a roughly annual basis (see {\tt http://www.sdss.org/science/index.html}).

This paper is based on two talks that I gave at the {\it New Era in Cosmology}
conference.  Section 2 reviews some of the recent SDSS results on 
galaxies and large scale structure; the topics mentioned are some of 
the ones that I have found interesting myself, and they by no means
constitute a comprehensive list.  All of these results are 
published or available on {\tt astro-ph}, so my summaries are brief
and do not include figures.  In Section 3, I discuss what we can hope
to learn from studying galaxy clustering in the SDSS and the 2dFGRS, with 
focus on the Halo Occupation Distribution as a way of thinking about the 
relation between galaxies and dark matter.  This discussion is based on 
collaborative work with Andreas Berlind, Zheng Zheng, Jeremy Tinker, and 
others.

\section{A Review of Recent Results}

\noindent
\subsection{Substructure in the Milky Way}

The first science results from the SDSS that really surprised me were
the discoveries, made independently by Yanny et al.\ (2000) using
A-colored stars and by Ivezi{\' c} et al.\ (2000) using RR Lyrae candidates,
of coherent, unbound structures in the
Milky Way's stellar halo, stretching across tens of degrees.
The idea that the stellar halo might be built by mergers of dwarf
galaxies is an old one (Searle \& Zinn 1978), and much of the recent theoretical
modeling has focused on detecting fossil substructure through
phase space studies of the local stellar distribution
(e.g., Johnston et al.\ 1995; Helmi \& White 1999; Helmi \& de Zeeuw 2000).  
Even in the absence of kinematic data, the SDSS is a powerful tool
for detecting substructure in the outer halo because multi-color imaging
allows the definition of samples of stars that are approximately standard
candles.  The two structures found by Yanny et al.\ (2000), in an area
$\sim 1\%$ of the sky, may both be associated with
the tidal stream of the Sagittarius dwarf galaxy (Ibata et al.\ 2001).
However, a recent study using F-stars (Newberg et al.\ 2001) appears
to show several more substructures, and no clear indication that there
is a smooth underlying halo at all.
Extending a model originally 
developed to investigate the dwarf satellite problem,
Bullock, Kravtsov, \& Weinberg (2001) showed that the population of
disrupted dwarfs expected in the CDM cosmological scenario could naturally
account for the entire stellar halo.  If this model is right, then
the SDSS should reveal ubiquitous substructure in the outer halo, where
orbital times are long and the number of discrete streams is relatively small.
In any event, the SDSS imaging survey will answer 
fundamental questions about the origin of the Milky Way's stellar
halo, and perhaps about the amount of power on sub-galactic scales
in the primordial fluctuation spectrum.

\subsection{Milky Way Satellites}

With multi-color data, one can define optimal filters to
maximize the contrast between an object in the Milky Way or Local Group
and the foreground and background
stellar distributions.  Odenkirchen et al.\ (2001ab) have developed
this technique and applied it to great effect in studies of the
globular cluster Pal 5 and the dwarf spheroidal Draco.
Pal 5 shows two well defined tidal tails that contain $\sim 1/3$ of
the cluster's stars, demonstrating that the cluster is subject to
heavy mass loss.  The orientation of the tails reveals the projected
direction of the cluster's orbit, and peaks within the tails may
be a signature of disk shocking events.  Draco, on the other hand,
shows no sign of tidal extensions even at surface densities
$\sim 10^{-3}$ of the central value, demonstrating that it is a bound,
equilibrium system, and justifying standard kinematic estimates
that yield a high mass-to-light ratio, $M/L_i \sim 100-150$.
As the SDSS covers more sky, we will get a more complete census of
which objects are being tidally destroyed and which are still
holding themselves together.  Tidal tails and tidal streams
may also provide valuable constraints on the radial profile, shape, and
clumpiness of the Milky Way's dark halo potential
(see, e.g., Johnston et al.\ 1999).

\subsection{The Galaxy Luminosity Function}

Blanton et al.\ (2001) measured the galaxy luminosity function 
in $u$, $g$, $r$, $i$, and $z$ 
using a sample of 11,275 galaxies observed during SDSS commissioning
observations.  The large sample size,
accurate photometry, and use of the Petrosian (1976) system for defining
galaxy magnitudes yield small statistical errors and excellent control
of systematic effects.  This analysis confirms and quantifies 
previous indications
that the galaxy luminosity function varies systematically with
surface brightness, color, and morphology; the first correlation
implies that low surface brightness galaxies make only a small
contribution to the mean luminosity density of the universe.
The measured luminosity density exceeds the $R$-band estimate from
the Las Campanas Redshift Survey (Lin et al.\ 1996)
by a factor of two, and  Blanton et al.\ show that this
difference arises from the isophotal magnitude definition adopted by the LCRS,
which misses light in the outer parts of galaxies that have intrinsically
low or cosmologically dimmed surface brightness.

Norberg et al.\ (2001b) demonstrate convincingly that the high-precision
measurement of the $b_J$-band luminosity function from the 2dFGRS is in
good agreement with the luminosity function measured by the SDSS.
Two main factors caused Blanton et al.\ (2001) to reach a contrary conclusion:
they used an inaccurate conversion from SDSS bands to $b_J$,
and their maximum likelihood method effectively estimates the luminosity 
function at a redshift $z \approx 0.1-0.15$,
while the method used by Norberg et al.\ (2001b) implicitly
corrects for evolution to derive the $z=0$ luminosity function.
Using the SDSS early release data, Norberg et al.\ (2001b) demonstrate
excellent agreement in the mean between SDSS and 2dFGRS galaxy magnitudes
and redshifts, and they confirm earlier estimates of the completeness and
stellar contamination of the 2dFGRS input catalog.
While the optical luminosity function estimates are now in good agreement,
there remains a puzzling discrepancy pointed out by Wright (2001) between
the luminosity density found in the optical bands by the SDSS
and the estimates of the $K_s$-band luminosity density from the
2dFGRS and 2MASS data (Cole et al.\ 2001;
see also Kochanek et al.\ 2001).  
Norberg et al.\ (2001b) argue that the discrepancy
is probably dominated by large scale structure fluctuations in the
area used to normalize the $K_s$-band luminosity function.

\subsection{Properties of Early Type Galaxies}

The SDSS data are ideal for studying the correlations of galaxy properties,
since the photometric and spectroscopic reduction pipelines measure many
quantities automatically and one can create large samples with well
understood selection effects.  The main effort to date in this area
is the comprehensive study of a sample of 9000 early type galaxies by 
Bernardi et al. (2001).  They determine the fundamental plane in 
the $g$, $r$, $i$, and $z$ bands, and they measure bivariate correlations 
among luminosity, size, velocity dispersion, color, mass-to-light ratios,
and spectral indices.  The large sample size and high precision 
allow examination of relatively subtle effects, such as a slight
difference in the fundamental plane of ``field'' and ``cluster'' galaxies.
The evolution of the fundamental plane over the redshift range of the
sample (which extends to $z\approx 0.2$) is consistent with passive
evolution of old stellar populations.  

\subsection{Galaxy-Galaxy Lensing}

One of the most dramatic breakthroughs from the SDSS has been the measurement
of galaxy mass profiles and the galaxy-mass correlation function via
galaxy-galaxy weak lensing.  Systematic effects are much easier to 
control for galaxy-galaxy lensing than for cosmic shear measurements
because image distortion is measured perpendicular to the radial separation
vector, which has a different orientation for each foreground-background pair.
The large area gives the SDSS great statistical power despite its
rather shallow imaging (by weak lensing standards).

Fischer et al.'s (2000) analysis of two nights of SDSS imaging
data (225 deg$^2$) was at the time the clearest detection of a 
galaxy-galaxy lensing signal, with extended shear profiles 
(to $r \ga 250 h^{-1}\;$kpc) offering direct evidence for the extended
dark matter halos expected in standard models of galaxy formation.  
McKay et al.\ (2001) have since analyzed a sample of
$\sim 3,600,000$ source galaxies around $\sim 35,000$ foreground
galaxies in the spectroscopic sample, measuring the galaxy-mass
correlation function and its dependence on galaxy luminosity,
morphology, and environment.
They find that the mass within an aperture of $260 h^{-1}\;$kpc
scales linearly with galaxy luminosity, that the excess mass density
around galaxies in high density regions remains positive
to $r \sim 1 h^{-1}\;$Mpc while that around isolated galaxies
is undetectable beyond $\sim 300 h^{-1}\;$kpc, and that early
type galaxies have a higher amplitude galaxy-mass correlation function,
in part because of their preferential location in group environments.
Guzik \& Seljak (2002) have modeled the McKay et al. results to
infer that $L_*$ galaxies have virial masses 
$M\sim (5-10)\times 10^{11}h^{-1}M_\odot$, implying that 
a large fraction of the baryons within the virial radius of an $L_*$
galaxy halo end up as stars in the central galaxy.
By comparing to the Tully-Fisher relation, Seljak (2002) concludes
that circular velocities at the halo virial radius are typically a
factor $\sim 1.7-1.8$ below the values measured at the galaxy optical radius,
and in reasonably good agreement with predictions based on CDM halo profiles.

\subsection{Angular Galaxy Clustering}

Early efforts to study galaxy clustering with the SDSS have focused
on the analysis of a $2.5^\circ \times 90^\circ$ stripe of imaging data
that has been closely examined and reduced multiple times.
Scranton et al.\ (2001) carried out an exhaustive analysis of possible
systematic effects associated with seeing variations,
stellar density, Galactic reddening, galaxy deblending, variations
across the imaging camera, and so forth, showing that they 
have no significant impact on measurements
of the angular correlation function at the obtainable level of
statistical precision.  These experiments demonstrate that 
star-galaxy separation in the SDSS imaging works extremely well
to $r\approx 22$.  Dodelson  et al.\ (2001) modeled the measurements
of the angular correlation function (Connolly et al.\ 2001) and
the angular power spectrum (Tegmark et al.\ 2001), 
to infer the 3-dimensional
clustering of galaxies.  Their results, $\sigma_8 \approx 0.8-0.9$,
$\Gamma \approx 0.15$, are consistent with those obtained by
Szalay et al.\ (2001) applying a different method, Karhunen-Loeve
parameter estimation, to the same galaxy catalog.
The statistical error bars from these analyses are not yet
competitive with the highest precision analyses of the
galaxy power spectrum (e.g., Percival et al.\ 2001),
but they provide reassuring evidence that any systematic biases
in the SDSS imaging data, and thus in the input to the redshift survey,
are well controlled.  Recently Szapudi et al.\ (2001) have 
analyzed the higher order angular moments of this data set, 
finding agreement with the hierarchical scalings and values of
skewness and kurtosis parameters predicted by $\Lambda$CDM models
that incorporate mild suppression of galaxies in high mass halos.
With the analysis tools now developed and tested
and the systematic issues apparently
well understood, the analyses of larger sky areas should soon yield
precise measurements of angular clustering over a wide
dynamic range.

\subsection{Clustering in the Redshift Survey}

Zehavi et al.\ (2001) carried out the first analysis of clustering
in the SDSS redshift survey, focusing on the real space correlation
function $\xi(r)$ 
and the pairwise velocity dispersion $\sigma_{12}(r)$
for different classes of galaxies,
with a sample similar in size and geometry to the LCRS.  
Galaxies in absolute magnitude bins centered on $M_*-1.5$, $M_*$, and
$M_*+1.5$ have parallel power-law correlation functions with 
slopes $\gamma \approx -1.8$, but their amplitudes are significantly
different, with $r_0\approx 4.7$, 6.3, and 7.4$h^{-1}\;$Mpc, 
respectively.  The correlation function of red galaxies is both
steeper and higher amplitude than that of blue galaxies.
The pairwise dispersion for the full sample
is $\sigma_{12} \approx 600 {\rm km\; s}^{-1}$ at $r\sim 1h^{-1}\;$Mpc,
but red galaxies have 
$\sigma_{12} \sim 700 {\rm km\; s}^{-1}$ and blue galaxies only
$\sigma_{12}\sim 400 {\rm km\; s}^{-1}$.  
The dependence of $\xi(r)$ on galaxy properties resembles that found
by Norberg et al.\ (2001ac) in the 2dFGRS (and in earlier studies
such as Guzzo et al.\ 1997), but there are significant differences
of detail.  The SDSS data show a steady trend of correlation strength
with luminosity, while Norberg et al.\ (2001a) find a transition from
weak dependence below $L_*$ to strong dependence above $L_*$.
Norberg et al.\ (2001c) find similar $\xi(r)$ slopes for galaxies
of different spectral types, while the correlation function of
blue galaxies in the SDSS is clearly shallower than that of red galaxies.
Analysis of larger SDSS samples should clarify the significance of 
these differences; the first could reflect the difference between
$r$-band and $b_J$-band selection, 
and the second could represent a difference between
color and spectral type as a basis for galaxy classification.
A consequence of the SDSS observing strategy (dictated largely by
the instruments themselves) is that the early redshift data
had a 2-dimensional slice geometry, making it difficult to study
the large scale power spectrum and statistics that require
contiguous 3-d volumes,
like void probabilities and topology.  That situation
is changing as the survey progresses, and first results on these
topics should emerge over the next several months.

\subsection{Random Optimistic Remarks}

The SDSS collaboration involves hundreds of scientists, with eleven
participating institutions on three continents.
With such a large and far-flung collaboration, we spend a lot of
energy just keeping ourselves organized.
I have just finished my term as the SDSS Scientific Publications Coordinator,
a position with one chief benefit: I was forced to pay attention 
as the scientific output of the SDSS grew from a trickle to a flood,
spreading rivulets into many different areas of astronomy.
This development has been exciting to watch, and I have learned
a lot of astronomy just by following it.
While there are certainly challenges of communication in a collaboration
this large, the process of going from data to science has worked,
in my opinion, remarkably well.  The ideal scenario is that each
scientific analysis draws on the collective expertise of a
very broad spectrum of astronomers; I have been delighted to see how often
we approach this ideal in practice.
The richness of the SDSS data is more than enough to keep us busy.
Indeed, while I am sure that we look enormous from the outside,
it is constantly evident from the inside that we don't have 
enough people to do all the science we would like to be doing.
That, of course, is one of many good reasons for publishing the data.

Before completely shifting gears, let me pause to congratulate the
members of the 2dF galaxy and quasar redshift surveys for
(a) obtaining more than 200,000 spectra, (b) publishing more than
100,000 spectra, and (c) writing a number of beautiful papers 
analyzing the results and implications.  All three of these are
great achievements.  While the SDSS and 2dF teams cannot help but
see themselves in competition every now and then, the benefits
to astronomy of having these independent data sets and independent
analyses are already very clear.

\section{What Can We Learn From Galaxy Clustering?}

The above question is one has been pondered by many people over several
decades.  Two developments that color recent considerations of this subject
are the extraordinary improvements in the
quantity and quality of the redshift survey data and the convergence of
the cosmological community on a ``standard'' model, $\Lambda$CDM,
that is supported by an impressive base of observational evidence.
A third development
that has deeply affected my own thinking
is the emergence of a new way of describing
galaxy bias, the Halo Occupation Distribution (HOD).  The HOD characterizes
the statistical relation between galaxies and mass in terms of the
probability distribution $P(N|M)$ that a halo of virial mass $M$ contains
$N$ galaxies, together with prescriptions that specify the relative spatial
and velocity distributions of galaxies and dark matter within these halos.
Note that ``halo'' here refers to a structure of typical overdensity
$\rho/\bar{\rho} \sim 200$, in approximate dynamical equilibrium;
higher density cores within a group or cluster are, in this description,
treated as substructure, and characterized only in a statistical sense.
Since different types of galaxies have different space densities and different
clustering properties, a given HOD applies to a specific class of galaxies,
e.g., red galaxies brighter than $L_*$, or late-type spirals with $r$
magnitudes $-17$ to $-19$.  The HOD framework has roots in early analytic
models that described galaxy clustering as a superposition of randomly
distributed clusters with specified profiles and a range of masses
(Neyman \& Scott 1952; Peebles 1974; McClelland \& Silk 1977).
A bevy of recent papers have shown that, when combined with numerical or 
analytic
models of the clustering of the halos themselves, the HOD is a powerful
tool for analytic and numerical calculations of clustering statistics, for
modeling observed clustering, and for characterizing the results
of semi-analytic or numerical studies of galaxy formation
(e.g., Jing, Mo, \& B\"orner 1998; Benson et al.\ 2000; Ma \& Fry 2000;
Peacock \& Smith 2000; Seljak 2000; Berlind \& Weinberg 2001; 
Marinoni \& Hudson 2001; 
Scoccimarro et al.\ 2001;
Yoshikawa et al.\ 2001; White, Hernquist, \& Springel 2001; 
Bullock, Wechsler, \& Somerville 2002).

My own interest in this approach was spurred largely by the paper of 
Benson et al.\ (2000), who discussed the clustering predictions of
their semi-analytic model of galaxy formation in these terms.
A forthcoming paper by Berlind et al.\ (in preparation; see also Berlind 2001)
compares the predictions of the Benson et al.\ semi-analytic formalism
to those of a large, smoothed particle hydrodynamics (SPH) simulation 
(Murali et al.\ 2001; Dav\'e et al., in preparation),
for the same cosmological model.  The agreement between the two 
approaches is remarkably good.  If we select galaxies above a specified
baryon mass threshold, chosen separately in the two calculations so that
the space densities of the populations are equal, then the mean occupation
$N_{\rm avg}(M)$ is a non-linear function of mass with three basic features:
a cutoff mass below which halos are not massive enough to host a galaxy
above the threshold, a low occupancy regime ($N_{\rm avg} \la 2$) in 
which the mean occupation grows slowly with increasing halo mass but
the average galaxy mass itself increases, and a high occupancy regime
in which $N_{\rm avg}(M)$ grows more steeply with mass, though the
growth is still sub-linear because larger, hotter
halos convert a smaller fraction of their baryons into galaxies.
In the low occupancy regime, the fluctuations about the mean are 
well below those of a Poisson distribution --- a halo that is supposed
to host one galaxy very rarely hosts two --- and the sub-Poisson nature
of these fluctuations has a crucial impact on some clustering statistics.
The HOD is strongly dependent on the age of the galaxies' stellar populations;
old galaxies like to live together in massive, high occupancy halos,
while young galaxies studiously avoid them.
The SPH simulations further show that the oldest, most massive galaxy in
a halo usually resides near the halo center and moves at close to the
center-of-mass velocity, while the remaining galaxies approximately
trace the spatial and velocity distribution of the halo's dark matter.
The agreement between the semi-analytic and SPH calculations, despite 
some clear differences in the way that they treat radiative cooling 
and feedback from star formation, suggests that the HOD emerges from
fairly robust physics that both methods do right, given their common 
assumptions.  Whether these assumptions hold in the real universe is,
of course, one of the things we hope to learn.

\begin{figure}
\plotone{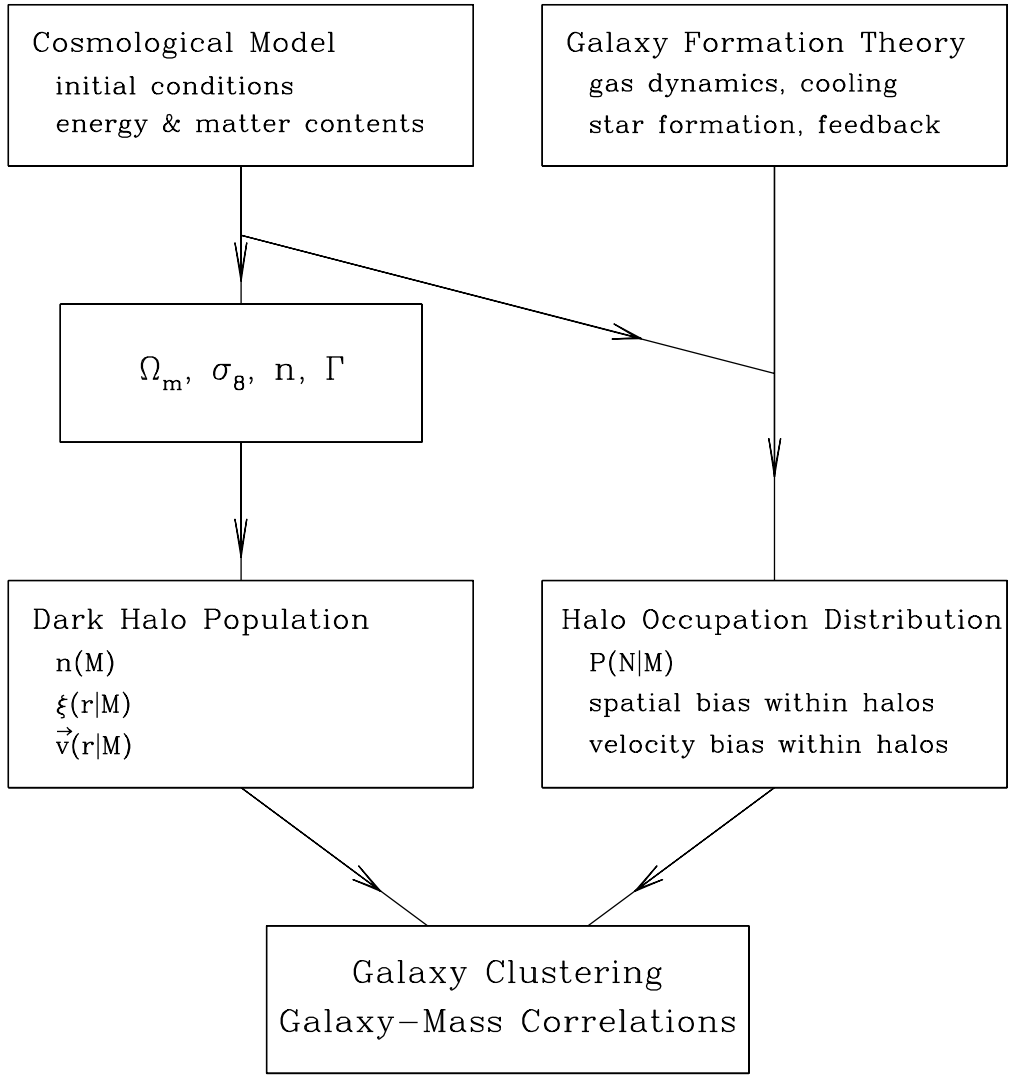}
\caption{The interplay between the cosmological model and the galaxy
formation theory in determining galaxy clustering and galaxy-mass 
correlations.}
\end{figure}

Figure 1 sketches the interplay between the ``Cosmological Model'' and
the ``Galaxy Formation Theory'' in determining galaxy clustering
(which I take to include the galaxy-mass correlations
measured by weak lensing).
The HOD approach suggests a nice division of labor between these two 
theoretical inputs.
The cosmological model, which specifies the initial conditions (e.g.,
scale-invariant fluctuations from inflation) and the matter and energy
contents (e.g., $\Omega_m$, $\Omega_b$, $\Omega_\nu$, $\Omega_\gamma$,
$\Omega_\Lambda$), determines the mass function, spatial correlations,
and velocity correlations of the dark halo population.  At our adopted
overdensity threshold $\rho/\bar{\rho} \sim 200$, these properties
of the halo population are determined almost entirely by gravity,
with no influence of complex gas physics.  I have inserted a box
in the path between cosmological model and dark halo population to
indicate that the only features of the cosmological model that really matter
in this context are $\Omega_m$, the fluctuation amplitude (represented
here by $\sigma_8$), and
the power spectrum shape (represented here by $n$ and $\Gamma$, though 
it could, of course, be more interesting).  Other features of the cosmological
model, such as the energy density and equation of state of the vacuum
component, may have an important impact on other observables or on the
{\it history} of matter clustering, but they have virtually no effect on
the halo population at $z=0$, if the shape of $P(k)$ and the present day 
value of $\sigma_8$ are held fixed.  The galaxy formation theory incorporates
the additional physical processes --- such as shock heating, radiative cooling,
conversion of cold gas into stars, and feedback of star formation on the
surrounding gas --- that are essential to producing distinct, dense, 
bound clumps of stars and cold gas.
It further specifies what aspects
of a galaxy's formation history determine its final mass, luminosity, diameter,
color, morphology, and so forth.  These physical processes operate in
the background provided by the evolving halo population, so the predicted
HOD depends on both the theory of galaxy formation and the assumed cosmological
model.

As a description of bias, the crowning virtue of the HOD is its completeness:
given a dark halo population and a fully specified HOD, one can predict the
value of any galaxy clustering statistic, on any scale, using analytic
approximations and/or numerical simulations.\footnote{There is one 
caveat here, namely the implicit assumption that a halo's
galaxy content depends, on average, only on its mass, and has no statistical
correlation with the halo's large scale environment.  This assumption is
adopted in ``merger tree'' formulations of the semi-analytic method, and
it is supported by the N-body experiments of Kauffmann \& Lemson (1999) 
and by the
results of the SPH simulation mentioned above, but it is not 
logically incontrovertible.}
Berlind \& Weinberg (2001) examined the influence of
the HOD on galaxy clustering and galaxy-mass correlations, for the
halo population of a $\Lambda$CDM N-body simulation.
We found that different clustering statistics, or even the same statistic
at different scales, are sensitive to different aspects of the HOD.
For example, at large scales $\xi_{\rm gg}(r)$ is proportional to the
mass correlation function $\xi_{\rm mm}(r)$, with a bias factor 
equal to the average of the halo bias factor $b_h(M)$ weighted by the halo
number density and the mean occupation $N_{\rm avg}(M)$.
On small scales, however, the explicit dependence on $\xi_{\rm mm}(r)$ 
disappears,
and $\xi_{\rm gg}(r)$ depends on the halo mass function, on the mean number
of pairs $\langle N(N-1)\rangle$ as a function of halo mass and virial radius,
and (to a lesser extent) on the internal bias between galaxy profiles and mass 
profiles.  Connecting these pieces into a power-law galaxy correlation function 
is a rather delicate balancing act, and the success of SPH simulations and 
semi-analytic models in reproducing the observed form 
of $\xi_{\rm gg}(r)$ given a $\Lambda$CDM
cosmology is entirely non-trivial; the reduced efficiency of galaxy formation
in high mass halos and the sub-Poisson fluctuations in low mass halos
are both crucial to this success.  Higher order correlation functions place
greater weight on the high mass end of the halo population and on
higher moments of $P(N|M)$.  
The void probability function, on the other
hand, depends mainly on the low mass cutoff of the HOD, which 
determines the probability of finding galaxies in the low mass halos
that populate large scale underdensities.
The pairwise velocity dispersion has distinct regimes much like 
$\xi_{\rm gg}(r)$, but it depends little on the low mass cutoff and
strongly on the relative occupation of high and low mass halos,
and the sub-Poisson fluctuations that depress $\xi_{\rm gg}(r)$ at
small scales {\it boost} the pairwise dispersion by forcing those
pairs that do exist at these separations to come from higher mass halos.
The pairwise dispersion can also be influenced by velocity bias of galaxies
within halos.  The group multiplicity function bears a quite direct
relation to the HOD, to such an extent that one can 
``read off'' $N_{\rm avg}(M)$
if $P(N|M)$ is reasonably narrow and one assumes an underlying halo
mass function $n(M)$.  Peacock \& Smith (2000) and Marinoni \& Hudson (2001)
have applied variants of this approach to observational data and
obtained results that agree rather well with the SPH and semi-analytic
predictions, assuming a $\Lambda$CDM halo mass function.

Berlind and I concluded that an empirical determination of the HOD should
be possible given high precision clustering measurements and the halo 
population of an assumed cosmological model.  This, at a minimum, is 
what we can expect to learn from galaxy clustering: the halo occupation
distributions of many different classes of galaxies, given a cosmological
model motivated by independent observations.  Because the HOD description
of bias is complete, these HODs encode everything that galaxy clustering
has to teach us about galaxy formation.  They encode it, moreover, in
a physically informative way, allowing detailed tests of theoretical 
predictions and providing rather specific guidance when these predictions 
fail.  If your theory of galaxy formation does almost 
everything right
but puts too many blue-ish S0 galaxies in $10^{13}-10^{14}M_\odot$ halos,
then you might have some ideas on how to fix it.

Can we have our cake and eat it too?  In more precise
words, if we find a combination of cosmological model and HOD that
matches all the galaxy clustering data, can we conclude that both are
correct, or might there be other combinations that are equally successful?

To decide whether cosmology and bias are degenerate with respect to
galaxy clustering, we must first know how changing the cosmology
alters the halo population.  This issue is the subject of a forthcoming
paper by Zheng et al., where we investigate the effect of changing 
$\Omega_m$ on its own, of changing $\Omega_m$ and $\sigma_8$ simultaneously
while maintaining ``cluster normalization'' 
($\sigma_8 \Omega_m^{0.5}=$constant), and of changing $\Omega_m$ and
$\sigma_8$ in concert with $n$ or $\Gamma$.
The impact of a pure $\Omega_m$ change is simple:
the halo mass scale $M_*$ shifts in proportion
to $\Omega_m$, pairwise velocities (at fixed $M/M_*$) are proportional to
$\Omega_m^{0.6}$, and halo clustering at fixed $M/M_*$ is unchanged.  
Cluster normalized changes to $\Omega_m$ and $\sigma_8$ keep the space
density of halos approximately constant
near $M\sim 5\times 10^{14}h^{-1}M_\odot$, and halo
clustering and pairwise velocities remain similar at fixed $M$.  
However, the shape of the halo mass function 
changes, with a decrease of $\Omega_m$ from 0.3 to 0.2 producing a
$\sim 30\%$ drop in the number of low mass halos.  One can preserve the 
shape of the mass function over a large dynamic range by changing $n$ or 
$\Gamma$, but the required changes are substantial --- e.g., masking
a decrease of $\Omega_m$ from 0.3 to 0.2 requires $\Delta n\approx 0.3$ or
$\Delta\Gamma\approx 0.15$.  These changes to the power spectrum significantly 
alter the halo clustering and halo velocities.  

The sensitivity of the halo population to the cosmological model
parameters is encouraging, because these changes cannot easily be masked
by changing the HOD.  For a pure $\Omega_m$ shift, one could keep the 
spatial clustering of galaxies the same by using the same HOD as a function 
of $M/M_*$, but the change would be detected by any dynamically sensitive
clustering statistic, like large scale redshift-space distortions, the 
pairwise velocity dispersion, the galaxy-mass correlation function, or
direct measurements of group and cluster masses.  Even velocity bias
within halos could not hide all of these changes.
A cluster-normalized change to $\sigma_8$ and $\Omega_m$ would require a 
change in galaxy occupation as a function of $M/M_*$ in order to 
maintain the galaxy space density and group multiplicity function, and 
this change would affect other measures of galaxy clustering.
A simultaneous change to the power spectrum shape that preserved the
halo mass function would change galaxy clustering by changing the clustering
of the halos themselves.

It remains to be seen just how well one can do quantitatively from realistic
observations.  The proof, ultimately, must await the pudding, but Zheng
has begun to investigate the question in a somewhat idealized context.
As a starting point, he takes clustering measures predicted by
a $\Lambda$CDM cosmology with the HOD derived from the SPH simulations,
calculated by a variety of analytic approximations.
He then changes the assumed cosmology, thus changing the halo mass
function and halo clustering, and he allows the HOD to change as well,
using a parametrized form that gives flexibility in
all of the essential features.  He finds the HOD that gives minimum $\chi^2$
for the original clustering ``measurements,'' which are assumed to have
10\% fractional uncertainties, and the value of $\Delta\chi^2$ 
for the best-fit HOD indicates
the acceptability of the cosmological model.  The preliminary results
from this exercise are encouraging.  For example, in the case of pure
$\Omega_m$ changes, the galaxy correlation function and group multiplicity
function constrain the HOD tightly enough that 
measurements of $\beta = \Omega_m^{0.6}/b$ or the pairwise velocity
dispersion impose useful constraints
on $\Omega_m$.  As the SDSS and 2dFGRS measurements take shape, we can
imagine taking a similar approach to the real data, albeit with careful
attention to the accuracy of the clustering approximations.
In terms of Figure~1, the surveys provide us with the entries in
the lowest box, and using them, we search for maximum likelihood 
solutions for the parameters in the second boxes on the
left and right hand sides.  Despite what might at first appear to be
a lot of freedom, the degeneracies appear to be limited, and we
can hope to do rather well.

Here, then, is my conjectural answer to the question posed in the section 
title:
we can learn the HOD of different classes of galaxies, gaining physical insight
into their origin, and we can separately determine $\Omega_m$ and the 
amplitude and shape of the linear theory power spectrum, from the largest
scales probed by the surveys (where perturbation theory describes the dark
matter and the HOD fixes the ``bias factors'' needed to connect galaxies to
mass) down to moderately non-linear scales (below which information about
the linear power spectrum may be effectively erased, at least as far as the
halo mass function and halo clustering are concerned).  We can also test for
any departures from Gaussian primordial fluctuations.
We get these cosmological constraints {\it without} relying on a detailed
theory of galaxy formation, only on the basic tenet that the HOD formulation
itself is valid.  While we might be wary of relying on conclusions that involve
complicated corrections for galaxy bias, the observed dependence of 
clustering on galaxy type allows powerful cross-checks.  
When we analyze different classes of galaxies, we should derive
different HODs, but we should 
always reach the same conclusions about the underlying cosmological model.
If we do, then we have good reason to think that we are doing things right.

Given all the other methods that can constrain cosmology with tracers
that are less physically complicated, one might wonder what galaxy clustering
and galaxy-mass correlations have to contribute to cosmological tests, beyond
a reassuring consistency check.  After all, how important is the second decimal
place on $\Omega_m$?  While I hear variants of this question often,
I think it is a red herring,
and that we should be relentless in our efforts to squeeze as much cosmological
information as possible out of galaxy redshift surveys.  Even if we assume
that there will be no major conceptual adjustments to the current leading
model, there are at least two fundamental issues on which precision 
measurements from galaxy clustering can play a critical role: 
the contribution (if any) of gravity waves to CMB anisotropy,
and the equation of state of dark energy.  The first can be addressed by 
a precise comparison between the CMB fluctuation amplitude and the present
day amplitude of matter fluctuations.
Evidence for or against gravity waves would take us much further in
understanding the origin of density fluctuations, and perhaps even to 
understanding the mechanism (inflation, colliding branes, ...) that accounts
for the size and homogeneity of the universe.  Galaxy clustering has no
sensitivity to the equation of state on its own, but the sensitivity of
other tests depends crucially on precise knowledge of $\Omega_m$, where
the combination of galaxy clustering and galaxy-galaxy lensing may ultimately
provide the best constraints.  Precise knowledge of today's fluctuation 
amplitude is also essential to some tests for the equation of state and
its time dependence (see, e.g., the discussion of Kujat et al.\ 2001).

Constraining gravity waves, the dark energy equation of state, and neutrino
masses are concrete goals that we can set for cosmological applications of
galaxy clustering.  But we should not assume that the simplest model consistent
with the current data (which already contains at least one very surprising 
element) will remain consistent with improving observations.  
A break in the inflationary
fluctuation spectrum, a relativistic background inconsistent with standard
neutrino physics, a baryon density inconsistent with big bang nucleosynthesis, 
a small admixture of non-Gaussian or isocurvature fluctuations --- all of these
are departures from the standard model whose quantitative impact would be 
subtle but whose physical implications would be profound.  What we will
learn from the 2dF and SDSS galaxy surveys depends in large part on what
the universe has to teach us, and that is something we cannot yet know.  
Finding out is an exciting task for the New Era in Cosmology.

\bigskip
I am grateful to my numerous colleagues in the SDSS for producing the 
exciting results that I have recapitulated in \S 2, and for
their efforts and progress in producing a data set that warrants the
theoretical musings in \S 3.  I thank my collaborators on the work
discussed in \S 3, especially Andreas Berlind, Zheng Zheng, and Jeremy
Tinker, whose contributions to the ideas and to the results have been central.
I thank the NSF for its support of this research program and the 
Institute for Advanced Study and the Ambrose Monell Foundation for
hospitality and support during the recent phases of this work.
More details about the SDSS, including links to the Early Data Release,
an ever-growing list of scientific publications based on the SDSS data,
and a list of the many participating institutions and funding agencies
that have made the survey possible, can be found at the official SDSS
web site, {\tt http://www.sdss.org}.

\end{document}